# COMPENDIUM OF FRONT-END ELECTRONICS

F. MESSI
Division of Nuclear Physics, Lund University and
European Spallation Source ERIC
Lund, Sweden
Email: francesco.messi@nuclear.lu.se

**Abstract**

Our world is changing fast. On one hand, technological developments provide us with new and powerful electronics devices on almost a weekly basis. On the other hand, the end-user of these electronics is now rarely required to actually configure the devices, as everything is automated. This "de-empowerment" of the end-user may be detrimental to the scientific community. This is because the unique conditions in which a scientific measurement is performed require the end-user to have unrestricted access to every variable of the experimental setup. In this compendium, a general overview of some popular electronic modules is presented, and the different characteristics the end-user should take into consideration when buying/designing electronics systems for a new experimental application are discussed. [1]

## 1. INTRODUCTION

Modern detectors provide as output electrical signals that encode the information obtained during measurements. These signals may be of very different nature depending upon the detector used. They may take the form of a current, a charge or a potential difference (dpp), to name few. For example, a photomultiplier tube (PMT) provides a signal in the form of current proportional to the amount of scintillation light collected on its photocathode. Regardless of the detector, electronic modules are required to process these signals and to decode the information contained within them. In the example of a scintillator instrumented with a PMT, reading the output current signal over a known resistance will provide a voltage signal (dpp pulse) while determining its amplitude will give a measurement of the charge deposited in the detector. In this way, instrumentation electronics are a key component of any experimental setup.

Depending upon the information required from the detector, a precise chain of signal-processing electronics needs to be defined. This electronics chain can be divided into blocks according to the standard function required of them. In the example of Figure 1, the electronics chain is divided into two blocks, a Front-End Electronics (FEE) block and a Read-Out Electronics (ROE) block. FEE is defined as the electronics located immediately downstream of the detector. Usually, dealing with analog signals, FEE are the most delicate part of the electronics chain. Consider first the concept of electronic noise, the signal-to-noise ratio (S:N) is crucial to analog signals. Noise is a random fluctuation of the voltage level and can affect the signal in different ways at different times. As an example, in Figure 2(a), the same pulse is recorded at four different times using a noisy electronic channel. As can be see, a measurement of the signal, either the integration of the pulse (charge) or the crossing of a fixed threshold value by the pulse (time), will provide four different results. Since analog signals carry all of the information obtained during the measurement in one single pulse, any noise perturbation may very easily compromise the readout.

---

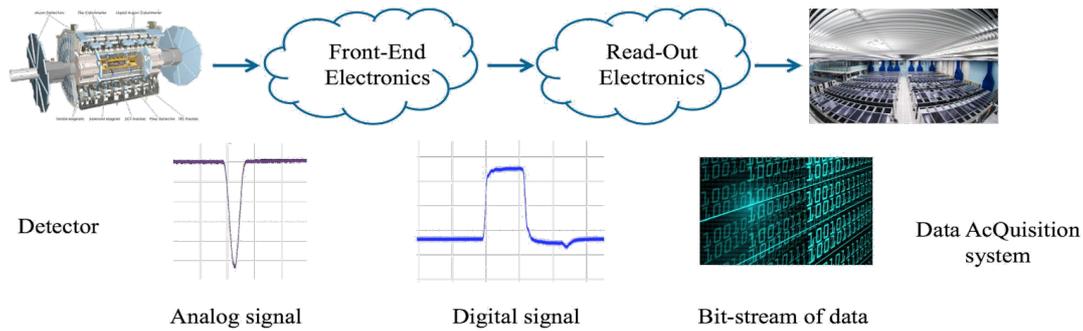

*Figure 1: Detectors of any experimental setup produce analog electrical signals. These analog signals are converted into digital signals by the Front-End Electronics (FEE). Read-Out Electronics (ROE) compress the information carried by the resulting digital signals into a bit-stream of data to be stored on a computer.*

In terms of signal integrity, a "digital signal" is more insensitive to noise than an "analog signal". A digital signal is, in fact, an analog signal which can have only two voltage levels. One level is assigned digital information "0" (*zero* or *low*) and the other level is assigned digital information "1" (*one* or *high*). The two levels are nonadjacent regions (the coloured bands in Figure 2(b)). The logical information contained within a digital pulse at a specific time depends whether the voltage level is in one region or in the other. Since each region is of the order of hundreds of millivolts, a noise perturbation needs to be substantially high to compromise the information carried by the pulse.

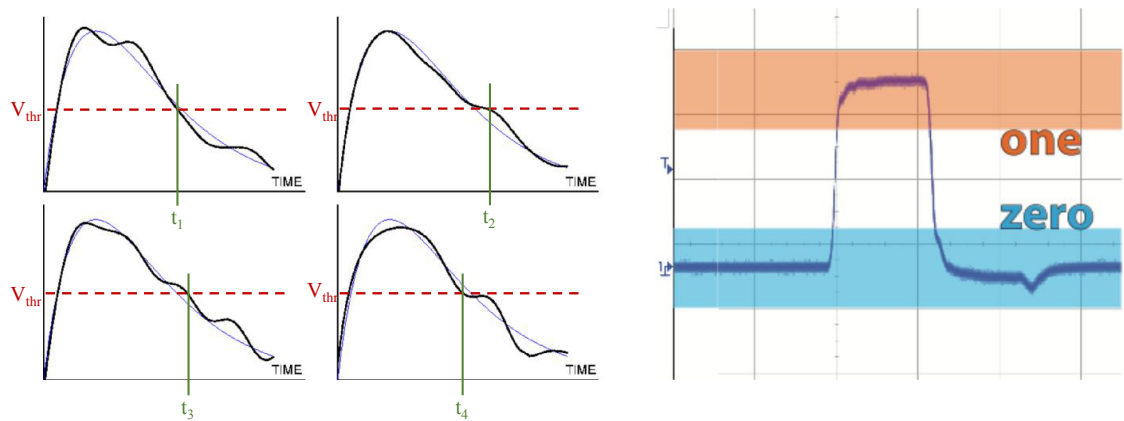

*(a) Signal plus noise (black) at four different times. The noiseless signal is superimposed for comparison (blue). The signal-to-noise ratio is about 20. Figure from [1], edit by the author.*

*(b) A digital signal can have only two voltage levels. The first is associated with digital information "zero" (blue band) and the second to digital "one" (orange band).*

*Figure 2: Analog signal (left) and digital signal (right). Analog signal are extremely sensitive to noise perturbations. The same pulse will cross a fixed threshold at four different times when noise is present.*

The purpose of the FEE is to:

1.  decouple the detector from the rest of the electronics system and match the impedances;
2.  acquire the electrical signal from the detector and prepare the information (shape the signal) for the ROE;
3.  digitise the analog signal for the Data AcQuisition system (DAQ).

## 1.1 DECOUPLING AND IMPEDANCE MATCHING

The main task of the FEE is to decouple the detector from the rest of the electronics chain and to match the impedances between the two. This is important to avoid damage both to the detector and to the electronics chain (such as from high-voltage discharges), and to avoid noise contamination in both directions. In fact, any spark resulting in a large transient electric signal from the detector may damage the electronics. For example, a broken wire from a wire chamber may short the connected digitiser to the high-voltage, thus damaging the electronics module and destroying the input stage of that channel. It is also important to protect the detector from any unwanted signal feeding back from the electronics. The majority of modern electronics modules are conceived for a 50 Ohm impedance system. This in general is not the case for the detector in use. If a mismatch in impedance is present, signal reflections may occur and/or an unwanted current may flow back into the detector. Both effects may introduce perturbations into the detector itself. For example, high-frequency noise on one channel may generate an oscillation on the ground plane of the detector, thus compromising all the channels. Moreover, if the impedance mismatch is very large, too much current may flow out of the detector resulting in damage to the detector itself.

There are several ways to decouple a detector. Very often, a solution is already integrated into the detector itself and not user-accessible. The simplest way to decouple a detector is via a resistor, a capacitor, or a combination of the two. This is known as "passive decoupling". Sometimes active decoupling is necessary, in order to preserve the integrity of the signal and/or to mach the impedances. In this case, an analog buffer or an operational amplifier (OpAmp) may be used. The decoupling of a system is unique and a specific study needs to be performed for each individual setup. It is responsibility of the user to check that the necessary decoupling is present and correct for the measurement to be performed.

## 1.2 ACQUISITION AND SHAPING OF THE SIGNAL

Signals from a detector are usually short current pulses *i(t)* whose time scale may vary from few *ps* till several hundreds of *µs*, depending upon the detector technology. These pulses may be invisible to the ROE. In fact, any electronics module needs a minimum amount of time to recognise an input signal and may miss a signal if it is too fast. Further, the pulse can be too small with respect to random noise or a slow discharge pulse and thus lie out of the sensitive range of the input stage of the applied electronics.

The shape of a pulse can be modified without affecting the information it carries. For example, a signal can be stretched in time, conserving its integral or maintaining it proportional to the original. In this way, a fast pulse can be smoothed and acknowledged by the ROE without perturbing the energy measure provided. These devices are known as "shapers". The simplest shaper consists of a series of CR-RC circuits (see Figure 4).

When the rate at the detector is high, a subsequent pulse can arrive before the previous pulse has been processed. If the second pulse overlaps with the previous pulse, this is known as "pileup". In this case, the electronics may not recognise the second pulse and its information may be lost (for example, to a discriminator) or erroneously added to the previous pulse (for example, by a Charge-to-Digital Converter, QDC). By constraining the pulse width, this effect can be limited and the second pulse can be fully resolved (see Figure 3).

Optimum pulse shaping generally depends on the particular application in question.

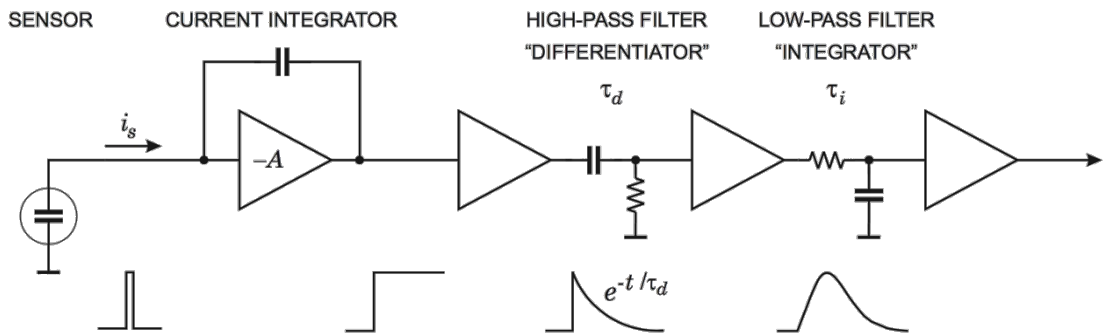

*Figure 4: Components of a pulse-shaping system. The signal current from the sensor, $i_s$, is integrated to form a step impulse with a long decay. A subsequent high-pass filter ("differentiator") limits the pulse width and the low-pass filter ("integrator") draws out the rise time to form a cusp-less pulse with a smooth transition. Figure from [1].*

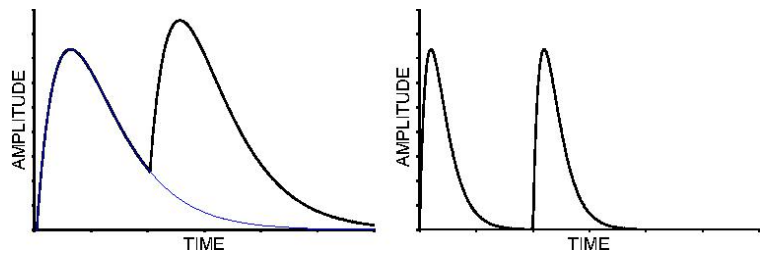

*Figure 3: Amplitude pileup occurs when two pulses overlap (left). Reducing the shaping time allows the first pulse to return to the voltage baseline before the second pulse arrives. Figure from [1].*

## 1.3  DIGITISING: FROM THE ANALOG TO THE DIGITAL WORLD

The ultimate task of the FEE is to prepare the information provided by the detector for the ROE or for the DAQ itself. This means the analog signal from the detector must be converted into a digital signal that can be passed on to a DAQ. In other words, the purpose of the FEE is to translate the information from the analog world to the digital world.

As mentioned before, a digital signal is an analog signal which can have only two voltage levels. Several different "logical families" exist, according to the different voltages assigned to the digital information. In Table 1, an overview is presented together with the respective corresponding voltages levels.

Conversion between the analog and digital world is performed with three different kinds of modules:

1. Discriminators: determine if a signal exceeds a given threshold.
2. Time-to-Digital Converters (TDCs): convert the time at which a signal is recorded into a digital number.
3. Analog-to-Digital Converters (ADCs): convert the analog value of a signal into a digital number. In particular, a Voltage-to-Digital Converter (VDC) converts an instantaneous voltage level, while a Charge-to-Digital Converter (QDC) the integrated charge.

| Family | | One (V) | Zero (V) |
|---|---|---|---|
| TTL | | 2 to 5 | 0 to 0.8 |
| NIM | | 3 to 12 | 2 to 1.5 |
| ECL | | 0.81 to 1.13 | 1.95 to 1.48 |
| LVDS | P | 1.27 to 2.40 | 0.92 to 1.12 |
| | N | 0.92 to 1.12 | 1.27 to 2.40 |

*Table 1: Families of digital signals. Not all families are compatible with each other. Intermediate modules that correctly translate from one family to another must be used to enable level transition.*

Any other more complex module is generally a combination or an evolution of these three basic models.

## 4. BASIC EXAMPLE: PHOTOMULTIPLIER TUBE (PMT)

Since the basic principles of FEE apply to any detector setup, a simple detector setup will be used for illustrative purposes in this paper — a scintillator coupled to a PMT.

Fig. 5 shows a scheme of three possible scenarios for the read-out of such a detector:

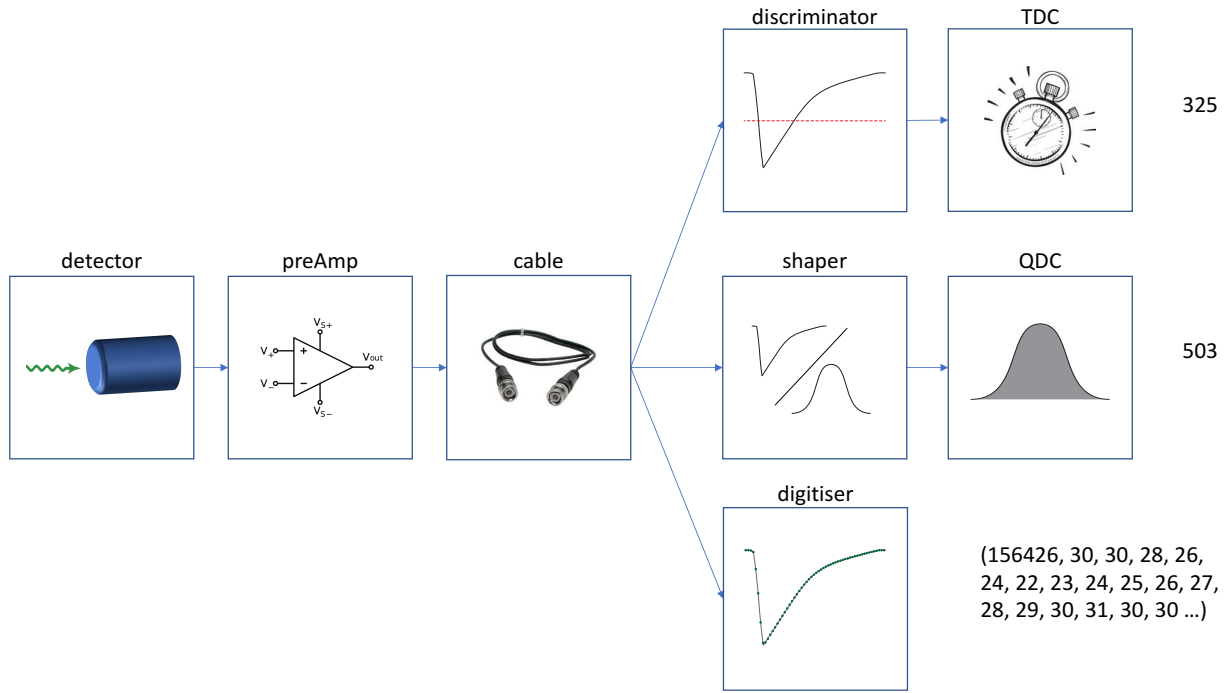

*Figure 5: Example of detector read-out. Three different electronics chain presented for the read-out of a scintillating detector coupled to a PMT.*

In the top case, a time measurement is performed. The signal is passed to a discriminator and a TDC is used to digitize time information. In the middle case, an energy measurement is performed. The signal is passed to a shaper circuit and a QDC is used to digitize the energy deposited in the detector. In the bottom case, the signal is passed to a digitiser. The entire pulse is digitised and, if the digital "image" of the pulse is accurate, all of the information is saved. Subsequent analysis can be performed off-line on the recorded data.

Two important components are common to all the scenarios. One is the first stage of the FEE: the pre-Amplifier (preAmp). The preAmp serves two purposes. The first is to decouple the detector to the rest of the electronics chain and match the impedance of the PMT to the impedance of the cable. The second is to amplify the signal from the detector (that in general can be very small) and to drive the transmission line, or cable. This helps preserving the signal-to-noise ratio (S:N). The second is the cable itself. Cables are crucial for any setup. It is through the cable that the signal travels from one component to another. It is also through the cable that the majority of noise is introduced. It is important to choose the correct cable for the setup. Different cables have different impedances, dielectric coefficients and can drive different kinds of signals.[2]

## 2.1 POSITION-TIME MEASUREMENT, DISCRIMINATOR AND TDC

Position-sensitive detectors measure the passage of a particle and the time of the detection. In the example of a plastic scintillator coupled to a PMT, the passage of a particle results in an electric pulse. To produce a trigger, the electric pulse needs to be discriminated from electronics noise and/or from signals generated by other particles interacting with the detector. The time of the digital pulse can be measured using a TDC.

### 2.1.1 Discriminator

A discriminator is a module that compare two signals. One of them is a fixed voltage level, also known as the "discrimination threshold", while the other is the analog pulse from the detector. If the pulse exceeds the threshold, the output of the discriminator will be "one". If the pulse is under threshold, the output is "zero" (refer to Figure 6). To avoid loss of information, the discrimination threshold needs to be set *as low as possible* as required by the S:N. In fact, the information carried by any signal that does not pass the discriminator is lost forever. Many kinds of discriminators exist, such as leading edge / dual threshold / constant fraction, to name few. A detailed discussion of these various types of discriminators is beyond the scope of this paper. In general, the user must be aware of the type of discriminator used and how to correctly set it up. The data sheet for the module may be useful.

### 2.1.2 TDC

A TDC is a device used to digitize the detected time of an event. In general, it requires a shaped signal as input. This is often a digital signal with constant width and very fast rise time, such as that provided by a discriminator. The simplest way to think of a TDC is a chronometer: a start is provided and a counter is increased at regular intervals (e.g. with a clock). When a signal is provided to the channel, the counter stops increasing. The time information relative to the signal is the number stored i n the counter multiplied by the period of the clock.

$$T_{measure} = TDC \times CLK$$

Modern TDCs, also called MultiHit-TDCs, do not stop the counter when the signal is received, but rather store the value of the counter into memory and continue to count. In this way, multiple hits can be recorded in sequence, without losing any events due to read-out and reset operations. There are several ways to generate a TDC and an explanation of them is beyond the scope of this paper. [3]

---

[2] For example, a coaxial cable is meant for *single-ended signals*, while a twisted pair ribbon cable is meant for *differential signals*. A detailed discussion on transmission lines in general is out of the scope of this paper. More information can be found in [2] or [3].

[3] Recently, FPGA (Field Programmable Gate Array) technology has been used more and more for TDC development. More information can be found in [4].

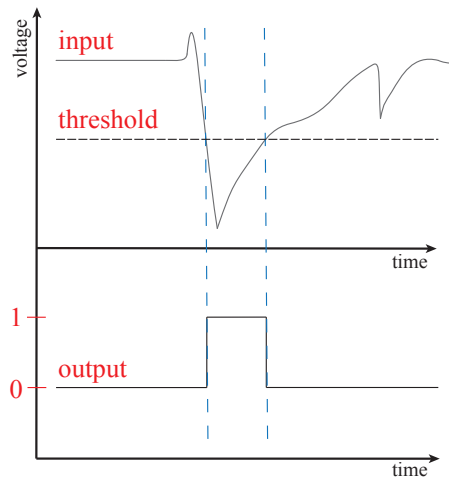

*Figure 6: Signals from a discriminator. On the top panel, an analog signal as function of time shown together with the constant discrimination threshold. On the bottom panel, the corresponding digital output signal form a discriminator.*

## 2.2 ENERGY MEASUREMENT, SHAPER AND QDC

Calorimeter detectors measure the energy deposited by a particle in the detector. In the example of a scintillator coupled to a PMT, the particle loses some or all its energy in the scintillator. The light emitted by the scintillating material, which is proportional to the deposited energy, is collected by the PMT. The resulting electrical current pulse has an integral that is proportional to the amount of detected light and thus the deposited energy. A QDC module will integrate the electric pulse, providing a digital word with a value proportional to the measured energy. Often, QDC modules require a signal with specific time and shape characteristics to correctly perform the integration. A shaper circuit is almost always needed (recall Section 1.2).

As in mathematics integral requires a close interval, so a conventional QDC needs a well-defined time interval for the integration of the pulse. This is provided with a digital signal called a "gate". The QDC will integrate the analog signal for the time period that the gate signal is "one". Analog delay lines are difficult to avoid in such a configuration as the signal to be measured needs to be placed within the gate. Recall that using cable is not as simple as it can seem. It is not always possible to obtain the necessary analog delay value to place the signal within the gate without compromising the integrity of the signal. The usage of digitiser modules or in general of a sampling ADC (sADC), is one way to reduce the amount of electronics modules in the setup and avoid analog delay lines.

## 2.3 DIGITISER

A digitiser is a module that allows for the conversion of analog signals into digital information that can be directly passed to a DAQ. As shown in Figure 7,the value of the analog signal is measured at regular steps. Those regular steps are known as the "sampling time". These samples, or relative digital amplitudes, can then be processed. In this way, a digital approximation to any pulse can be stored on a disk and the full measurement can be analysed in real time or off-line. In fact, once the pulse is reconstructed, the time of the interaction can be extrapolated as can the charge integration or the shape identification.

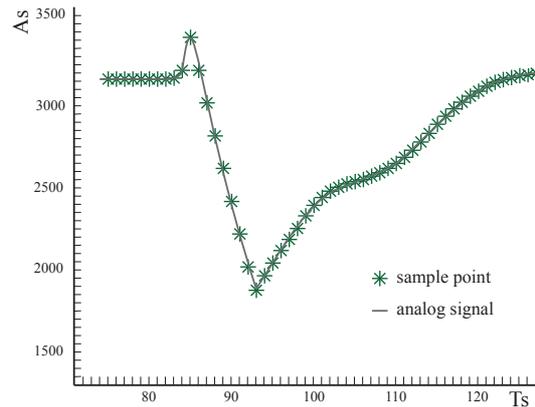

*Figure 7: Digitiser output. In grey the original analog signal from a PMT; in green, the sample points. The ADC channels (As) are plotted as a function of their sampling time (Ts).*

### 3. CHARACTERISTICS OF A DIGITISER

Modern electronics are everyday becoming faster and delivering better performance. Today, it is possible to perform all of the discrimination, shaping and digital conversion operations with a digitiser for the majority of applications. In fact, the digital representation of an analog pulse provided by a modern digitiser is so accurate that all the original information enclosed within the original analog signal can be reconstructed offline using mathematical algorithms. For each of the sample points, the digitiser provides a pair of numbers: the *sampling time* ($T_s$) and the ADC value of the sample ($A_s$). By plotting these variables against one other, the digitised pulse can be visualised. This is exactly the same process performed by a modern oscilloscope — the input signal is digitised and the sampled points are displayed as function of the time. From such a figure, the main characteristics of many electronics modules can be examined:

**Sampling Rate.** The Sampling Rate ($R_s$), expressed often in Gs/s *(giga samples per second)*, represents the constant time interval between two consecutive measurements. The higher the sampling rate, the more accurate the reconstruction of the signal. As a consequence, more memory is required to store the data. If the sampling rate is too low, features of the original signal may be lost, as shown in Figure 8(a). If the rate of the sampling is high enough [5], the original signal can be reconstructed without losing any of the information it carried.[4]

**Slew Rate.** The Slew Rate, expressed often in V/μs *(volts per microsecond)*, is defined as the maximum change of voltage per unit of time. It represents the number of volts one measure-ment can differ at maximum from the previous if the sampling rate is 1 μs. This is a very important characteristic! If the Slew Rate is not high enough, the digitiser will not be able to follow the signal. Regardless of the sampling rate, the signal will not be reconstructed correctly (see Figure 8(b)).[5]

---

[4] It applies also to: sADC.
[5] It applies also to: OpAmps, analog buffers, discriminators.

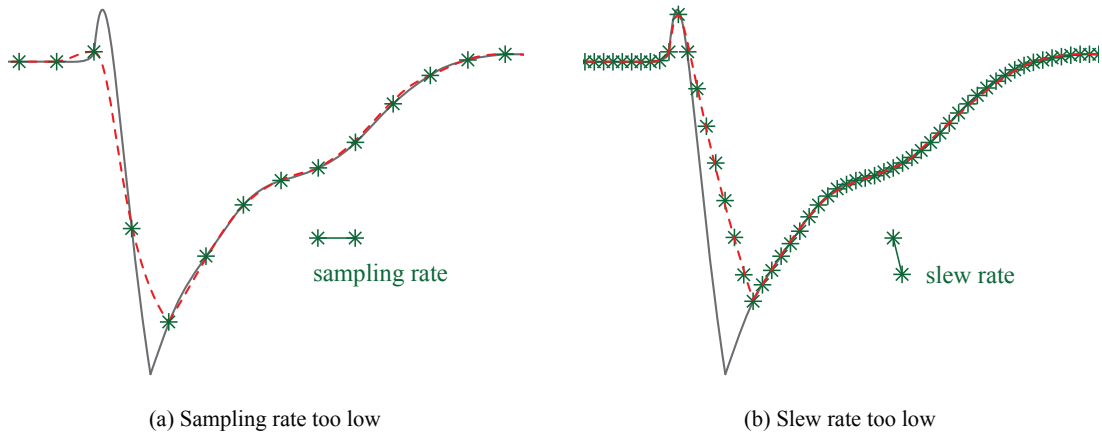

(a) Sampling rate too low       (b) Slew rate too low

*Figure 8: Non-optimal digitization of an analog pulse. In grey the original analog signal, in green the digitised samples and in red the signal reconstructed from the digitisation. If the Sampling Rate and the Slew Rate are not high enough, important features of the original signal may be lost.*

**Analog Bandwidth.** The analog bandwidth represents the range of frequencies for which the device performs linearly. Any signal can always be decomposed in its Fourier transform. The maximum frequency of this series must be included within the bandwidth. To avoid distortion, the bandwidth of the digitiser needs to be large enough for the detected signal. On the other hand, if the bandwidth is too large the amount of noise transmitted without increasing the signal will increase.[6]

**Input Range.** The input range is the maximum amplitude the digitiser can accept as input. If the input signal exceeds this, the digitiser will first saturate and, if the value continues to increase, may be damaged.[7]

**Reference Voltage and Analog Resolution.** Reference voltage and analog resolution are two characteristics linked to the digitisation process itself. A digital representation of an analog signal cannot have infinite precision. It would require a sampling process with an infinite sampling rate, slew rate and analog bandwidth. A real sampling (discretisation) process occurs with finite voltage range divided into a well-defined number of sub-intervals. The upper limit of the range is called the "reference voltage", while each interval of the range corresponds to an ADC channel ($A_s$). The analog resolution of a digitiser, expressed in the number of bits, is given by the voltage range divided the total number of ADC chan-nels. In Figure 7, the baseline of the signal is at channel ~3150. In binary, this is equal to *110001001110* and the digitiser used must have an analog resolution of at least 12 bits. Assuming the reference voltage range from -1 to +1 V, each ADC channel corresponds to 2 V / $2^{12}$ ch ~ 0.5 mV.[8]

**Time Resolution.** The time resolution is the depth of the TDC counter register. It is expressed in "number of bits" and provides the number of consecutive samples to which a unique time stamp can be assigned ($\#_s = 2^n$). In the majority of the cases, after a time $\#_s \times R_s$, the counter is reset without an intervention from the user.[9]

---

[6] It applies also to: OpAmps, analog buffers.
[7] It applies also to: any kind of electronics modules.
[8] It applies also to: ADCs, TDCs.
[9] It applies also to: TDCs.

**Memory.** While in principle it is possible to transfer all data as soon as they are measured, this is inconvenient and highly demanding on the hardware in question. Instead, data are often stored inside the module and transferred to the user (to the screen of the oscilloscope or to the DAQ running on a computer for example) in blocks. A large amount of memory may help in protecting against data loss during the transmission stage.[10]

4. SUMMARY

Modern electronics are becoming cheaper every day and simultaneously offering better and better performance. Today, it is possible to perform scientific experiments using commercially available electronics, without any need to develop custom modules for individual detectors. Nevertheless, it is very important to choose the correct FEE and to correctly configure it, according to the needs of the experiment to be performed. The FEE are the electronics located as close as possible to the detector. They serve to decouple the detector from the rest of the electronics chain and match the impedances, to shape the analog signal provided by the detector, and to convert this analog signal to a digital signal. Modern digitisers can recreate an accurate digital representation of the analog pulse from a detector. In the majority of cases, this representation is sufficient for all the information carried by the analog pulse can be correctly reconstructed.

---

[10] It applies also to: ADC and TDC boards.